\documentclass[prl,aps,superscriptaddress,amsmath,amssymb,showpacs,floatfix,twocolumn]{revtex4}
\usepackage[pdftitle={Non-contact dissipation on the surface of charge density wave materials}]{hyperref}
\usepackage{epsfig}
\usepackage{amsfonts}
\usepackage{amsmath}
\usepackage{natbib}
\usepackage{dcolumn}
\newcommand{\md}[1]{\left|#1\right|}
\newcommand{\FE}{\mathcal{F}}
\newcommand{\qq}{\mathbf{Q}}
\newcommand{\rr}{\mathbf{r}}

\begin{document}

\title{Non-contact dissipation on the surface of charge density wave materials}

\author{Franco Pellegrini}
\affiliation{SISSA, Via Bonomea 265, I-34136 Trieste, Italy}
\affiliation{CNR-IOM Democritos National Simulation Center, Via Bonomea 265, I-34136 Trieste, Italy}

\author{Giuseppe E. Santoro}
\affiliation{SISSA, Via Bonomea 265, I-34136 Trieste, Italy}
\affiliation{CNR-IOM Democritos National Simulation Center, Via Bonomea 265, I-34136 Trieste, Italy}
\affiliation{International Centre for Theoretical Physics (ICTP), P.O. Box 586, I-34014 Trieste, Italy}

\author{Erio Tosatti}
\affiliation{SISSA, Via Bonomea 265, I-34136 Trieste, Italy}
\affiliation{CNR-IOM Democritos National Simulation Center, Via Bonomea 265, I-34136 Trieste, Italy}
\affiliation{International Centre for Theoretical Physics (ICTP), P.O. Box 586, I-34014 Trieste, Italy}

\date{\today}

\begin{abstract} 
Bulk electrical dissipation caused by charge-density-wave (CDW)  depinning and sliding is a classic subject. 
We present a novel local, nanoscale mechanism describing the occurrence of mechanical dissipation 
peaks in the dynamics of an atomic force microscope tip oscillating above the surface of a CDW material. 
Local surface 2$\pi$ slips of the CDW phase are predicted to take place giving rise to mechanical hysteresis
and large dissipation at discrete tip surface distances. 
The results of our static and dynamic numerical simulations are believed to be relevant to recent experiments on NbSe$_2$;
other candidate systems in which similar effects should be observable are also discussed.
\end{abstract}


\maketitle

\section{How it all began: Gabriele Giuliani and CDW in the 70'} 

{\it Erio Tosatti} -- When I first met Gabriele --- here ``I'' is ET --- it was 1975, when he turned 
up in my office at the University of Rome,
introduced by my senior colleague and former mentor Franco Bassani.  Gabriele was then a 
young undergraduate student of Pisa's Scuola Normale (where I also came from) and was seeking outside
advisors for a thesis subject in modern condensed matter theory. My Rome colleague Mario Tosi,
who was a professor, accepted to serve as his formal external advisor, and so Gabriele started coming
back periodically to Rome, working on the general subject of the electron gas --- ``{\em Electron Gas}'' 
even became Gabriele's nickname, as far as I was concerned. The age gap between Gabriele and me 
was slightly less than ten years; we became friends and spent time together, talking physics, politics, 
and  everything else on our minds.  I had recently come back from Cambridge where I had worked with
Phil Anderson on a possible surface version of charge-density-waves (CDWs) instabilities of the electron gas
invented a decade earlier by Al Overhauser in the US. I was enthusiastic about the subject,
and ended up getting Gabriele interested too, and so it was that CDWs and Overhauser 
made their first entry in Gabriele's life.

\begin{figure}[!tb]\centering
\includegraphics[width=0.48\textwidth]{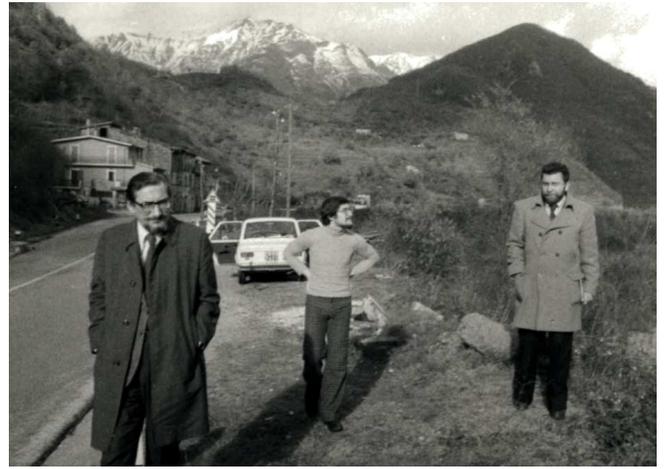}
\caption{\label{fig:system} Gabriele Giuliani (center) with Mario Tosi (left) and Franco Bassani (right)
on the Appennines during a trip from Rome to Gabriele's hometown, Ascoli Piceno (Autumn 1976). 
Photograph taken by Erio Tosatti.}
\end{figure}

In 1977 when, after completion of his degree in Pisa, I asked Gabriele to consider my newly founded 
group in Trieste, to which I had managed to attract colleagues of the caliber of young Michele Parrinello 
and of mature Mario Tosi. He liked it, and decided to join. In Trieste, we began working on one-dimensional CDWs and
similar systems \cite{Giuliani_Tosatti_Tosi_LNC76,Giuliani_Tosatti_NC78,Giuliani_Tosatti_Tosi_JPC79}.  
He worked hard and got far ahead of my rudimentary command of many body theory and related suggestions.
I recall for example one calculation where he was supposed to reproduce, with some supposedly clever approximation
I had cooked up, one exactly known result. I abused him abundantly because he was consistently failing 
to get the exact result by a factor 2 --- only to discover the hard way that he was right and the culprit was 
my beloved approximation.  In spite of all that, when one summer I met Al Overhauser in the US, he 
approached me quite enthusiastically about our otherwise universally ignored CDW paper \cite{Giuliani_Tosatti_NC78}, 
and declared that he would be delighted to welcome Gabriele to visit his group, possibly for a PhD curriculum
(a title that did not exist in Italy at the time).
And this is how Gabriele ended up at Purdue, in 1979, first as a postdoc with Al Overhauser, and then, after a further 
postdoc at Brown University, with a tenure. 
Our more than brotherly relationship continued uninterrupted to his very last few days, with visits, contacts, and
many many phone and Skype calls which he would never forget to make in connection with all kinds of occasions
or even without.  Besides Gabriele's friendship, his other main present has been his former student Giuseppe Santoro, 
who came to Trieste at his suggestion to become a close friend and collaborator to this day. 

{\it Giuseppe Santoro} -- It was 1987 --- a few years after he returned to Purdue from Brown --- that I joined Gabriele at Purdue, 
the ``I'' now  being GES. I had met him already in a couple of occasions in Pisa, were I was a student of the Scuola Normale Superiore
that Gabriele regularly visited during his trips to Italy, to visit his family in the beloved hometown of Ascoli Piceno
({\em Al\'e Ascoli} was, as I soon discovered, Gabriele's first screen upon log-in on any computer on earth). 
By that time --- we were in the Quantum Hall Effect era --- the focus of Gabriele research was the two-dimensional 
electron gas, and this was the initial interest of my first papers at Purdue. But I still cannot recall a single office day,
during my stay at Purdue from 1987 to 1991, in which Gabriele was not paying a visit to Al Overhauser: and
in Overhauser's office, a most sure topic of discussion was charge (and spin) density waves.  

It is fair to say that, since the early days, Gabriele and CDW (first) and two-dimensional electronic systems (later) remained
entangled together.  
%
It is thus fitting that this article, devoted to him and his legacy, should be on the CDWs of a peculiar two-dimensional layered
system, NbSe$_2$, on which we have recently come across.

\section{Introduction}
Charge-density-waves (CDWs) are static modulations of small amplitude and generally incommensurate
periodicities which occur in the electron density distribution and in the lattice positions of a variety of  
materials~\cite{Gruner_RMP88}. They may derive either by an exchange-driven instability of a metallic 
Fermi surface~\cite{Overhauser_PR68},  or by a lattice dynamical 
instability leading to a static periodic lattice distortion (PLD) which may equivalently be driven by electrons 
near Fermi~\cite{Peierls_55,Woll_PR62} or finally just by anharmonicity~\cite{Weber_PRL11}.  
A CDW superstructure, characterized by amplitude $\rho_0$ and phase $\phi$ relative to the underlying 
crystal lattice can be made to slide with transport of mass and charge and with energy dissipation under 
external perturbations and fields~\cite{Gruner_RMP88}. 

Phase slips in bulk CDWs/PLDs are involved in a variety of phenomena, including noise 
generation~\cite{Coppersmith_PRL90}, switching~\cite{Inui_PRB88}, current conversion at contacts~\cite{Maher_PRL92}, 
noise~\cite{Ong_PRL84,Gruner_PRL81} and more. While these phenomena are now classic knowledge, 
there is to date no parallel work addressing the possibility to mechanically provoke CDW phase slips 
at a chosen local point, see pictorial illustration in Fig.~\ref{fig:system}. 
In this work we describe a two-dimensional model showing how a localized CDW/PLD 
phase slip may be provoked by external action of an atomic force microscope (AFM) tip at an arbitrarily 
chosen point outside a surface.  
\begin{figure}[!tb]\centering
\includegraphics[width=0.48\textwidth]{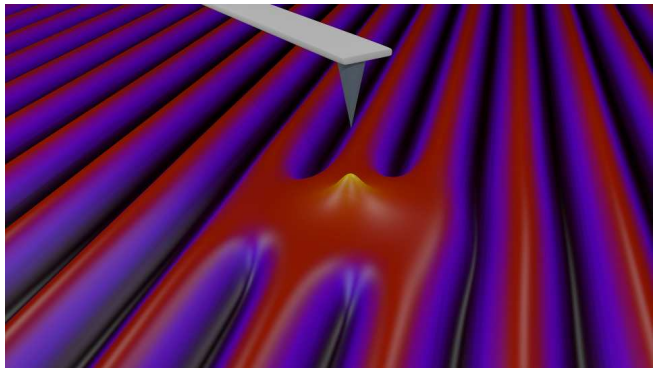}
\caption{\label{fig:system} Pictorial view of an AFM tip provoking phase slips over a surface CDW 
modulation.}
\end{figure}

The study of the microscopic mechanisms leading to energy dissipation and friction has very important theoretical
and practical implications. In recent years, experiments have started to single out the effects of microscopic 
probes in contact or near contact with different surfaces, and much theoretical effort has been devoted 
to the understanding of such experiments~\cite{Vanossi_RMP13}.
In particular, the minimally invasive non-contact experiments offer a chance to investigate delicate surface 
properties and promise to bring new insight on localized effects and their interaction with the bulk. 
The development of ultra-sensitive tools such as  the ``pendulum'' AFM~\cite{Stipe_PRL01,Gysin_RSI11} offers a chance 
to investigate more delicate and intimate substrate properties. Near a CDW material the tip oscillations may 
actuate, through van der Waals or electrostatic coupling,  an electronic and atomic movement in the surface 
right under the tip, amounting in this case to coupling to the CDW order parameter. 
Owing to the periodic nature of the CDW state, the coupled tip-CDW system has multiple solutions, characterized 
by a different winding number (a topological property) which differ by a local phase slip, and correspond 
to different energy branches. At the precise tip-surface distance where two branches cross, 
the system will jump from one to the other injecting a local 2$\pi$ phase slip, and the corresponding hysteresis 
cycle will reflect directly as a mechanical dissipation, persisting even at low tip oscillation frequencies. 

Recently, a non-contact atomic force microscopy (AFM) experiment~\cite{Langer_NatMat14} on a NbSe$_2$ sample has shown  
dissipation peaks appearing at specific heights from the surface and extending up to $2$ nm far from it. These peaks
were obtained with tips oscillating both parallel and perpendicular to the surface, and in a range of temperatures
compatible with the surface charge density wave (CDW) phase of the sample.
In this paper, a model is proposed explaining in detail the mechanism responsible for these peaks: the tip 
oscillations induce a charge perturbation in the surface right under the tip, but, due to the nature of the CDW order
parameter, multiple stable charge configurations exist characterized by different ``topological'' properties.
When the tip oscillates at distances corresponding to the crossover of this different manifolds, the system is 
not allowed to follow the energy minimum configuration, even at the low experimental frequencies of oscillation,
and this gives rise to a hysteresis loop for the tip, leading to an increase in the dissipation.

\section{The Ginzburg-Landau model} 
In the following, we will use the term CDW to indicate a periodic modulation of the charge density $\rho$,
irrespective of the process behind its generation. 
This modulation is described, in the unperturbed system and for the simplest form of CDW, 
as $\Delta\rho(\rr)=\rho_0\cos(\qq\cdot\rr+\phi_0)$, where $\rho_0$ is the intensity, $\lambda = 2\pi Q^{-1}$ the characteristic wavelength, and $\phi_0$ an initially constant phase, fixed by some far away agent.
Perturbations to CDWs have been studied extensively~\cite{Gruner_RMP88,Fukuyama_PRB78,Lee_PRB79,Tucker_PRB89}, but most studies are concerned either with uniform perturbations (e.g., the dynamics of a CDW under an external electric field) or 
point-like perturbations  (e.g., the static pinning of the CDW by defects), and often consider one-dimensional models, 
appropriate for quasi-one-dimensional materials, where the coherence length in the perpendicular directions is smaller 
than the atomic distance. 
Here we wish to study, instead, the effect of a localized perturbation represented by
a weakly interacting and slowly oscillating nano or mesoscopic sized probe hovering above the surface, acting on a 
length scale $\sigma$ similar to the CDW wavelength, $\sigma\sim 2\pi Q^{-1}$, 
and on a material where the coherence length is macroscopic in more than one dimension.

Starting from the standard Fukuyama-Lee-Rice model~\cite{Fukuyama_PRB78,Lee_PRB79} for CDW, the charge modulation is 
described as a classical elastic medium, through a Ginzburg-Landau (GL) theory. 
A complex space-dependent (and later, time-dependent) order parameter $\psi(\rr)=A(\rr)e^{i\phi(\rr)}$ will take into account 
both the amplitude degree of freedom $A$, as well as the phase $\phi$, in terms of which the charge density modulation 
is expressed as $\Delta\rho(\rr)=A(\rr)\cos(\qq\cdot\rr+\phi(\rr)) = \mathrm{Re} \left[\psi(\rr)e^{i\qq\cdot\rr} \right]$. 
The unperturbed system has $A(\rr)=\rho_0$ and $\phi(\rr)=\phi_0$, both constant. 
The GL free-energy functional, in absence of any external perturbation, will read:
\begin{equation} \label{eq:FE0}
\FE_0[\psi(\rr)]=\int \! \mathrm{d}\rr \, \left[-2f_0\md{\psi(\rr)}^2+f_0\md{\psi(\rr)}^4+\kappa\md{\nabla\psi(\rr)}^2\right]\;,
\end{equation}
where $f_0$ sets the energy scale and $\kappa$ accounts for the elastic energy cost.
If we now consider the effect of an external perturbation --- in our case, the AFM tip, generically described as a potential $V(\rr)$ 
coupling to the charge density modulation $\Delta\rho(\rr)$ ---, we will need to add to the GL free-energy functional
an extra term of the form:
\begin{equation} \label{eq:FEV}
\FE_V[\psi(\rr)] = \int \! \mathrm{d}\rr \, V(\rr) \, \Delta\rho(\rr) 
= \int \! \mathrm{d}\rr \, V(\rr) \, \mathrm{Re} \left[ \psi(\rr)e^{i\qq\cdot\rr} \right] \;.
\end{equation}
The literature \cite{Tucker_PRB89,Tutto_PRB85} has dealt extensively with the case in which $V(\rr)$ represents 
the potential due to impurities present in the sample, where it is appropriate to take $V(\rr)=\sum_i\delta(\rr-\rr_i)$,
since the typical scale in which the impurity potential acts is much smaller than $2\pi Q^{-1}$. 
In that case, phase-only oscillations --- with an essentially constant amplitude $A(\rr)$ --- are often enough to 
study the ground state of the system, resulting from the balance of elastic and potential energy, and
described by a phase-only GL functional of the form:
\begin{equation} \label{eq:FEphi}
\FE_{\phi}[\phi(\rr)]=\int \! \mathrm{d}\rr \, \left[\kappa\md{\nabla\phi(\rr)}^2+V(\rr)\rho_0\cos(\qq\cdot\rr+\phi(\rr))\right]  \;.
\end{equation}
Extremely localized impurity perturbations of this sort, however, only impose a likewise point-like constraint on the phase of
the order parameter, and cannot lead to a phase slip, in the absence of an external driver~\cite{Tucker_PRB89}.
To model an AFM tip, on the contrary, we should consider the case where $V(\rr)$ has a given specific shape
with a finite width $\sigma$ of the order of the wavelength $2\pi Q^{-1}$, and minimize the total GL free energy 
$\FE=\FE_0+\FE_V$, including the amplitude degree of freedom $A(\rr)$.
The fact that a phase-only functional $\FE_{\phi}$ is inadequate in describing this specific effect can be argued as follows. 
If we consider a purely one-dimensional model $\FE_{\phi}[\phi(x)]$, we would end-up with a linear 
behavior of $\phi(x)$ --- the solution of the Laplace equation in one-dimension --- in the regions where the potential is zero. 
Since we expect a decay of $\phi$ towards some constant $\phi_0$ far from the perturbation, 
this is a clearly unphysical result. 
But moving to a two-dimensional phase-only functional does not improve the situation very much.
Indeed, due to the nature of the phase, which is defined modulo $2\pi$, given some boundary conditions the solution
is not univocally defined unless the total variation of $\phi$ along the sample is also specified. 
Assuming the phase to have the unperturbed value $\phi_0$ far from the perturbation, we can define the integer 
\textit{winding number} $N$ of a solution as the integral
\begin{equation} \label{eq:WN}
N=\frac{1}{2\pi}\int\nabla\phi(x)\mathrm{d}x \;,
\end{equation}
taken along the CDW direction $\qq$ (with $N=0$ typically representing the unperturbed case).
Since any change in the winding number along the $\qq$ direction would extend to the whole sample, and unnaturally 
raise the energy of such a solution, to recover a physical result we definitely need to take into account the amplitude 
degree of freedom, which will allow for the presence of dislocations and local changes in the winding number.

For these reasons, we consider the full GL problem $\FE=\FE_0+\FE_V$ in two dimensions, 
with the complete complex order parameter $\psi(\rr)=A(\rr)e^{i\phi(\rr)}$, and in a subspace with a definite winding number.
The final result is expected to be similar to what previously considered in the wider context of phase 
slips~\cite{Tucker_PRB89} and more specifically in the case of localized phase slip centers~\cite{Maki_PLA95,Gorkov_ZETF84}. 
Namely, the local strain induced by the perturbation on the phase will reduce the order parameter amplitude,
to the point where a local phase slip event becomes possible. 
In more than one dimension, the boundary between areas with different winding number will be marked 
by \textit{vortices} of the phase.

From this preliminary analysis, the mechanism responsible for the dissipation peaks can be understood: as the 
tip approaches the surface, it encounters points where the energies of solutions with different winding number 
undergo a crossover. At these points the transition between manifolds is not straightforward, due to the mechanism 
required to create the vortices; therefore the tip oscillations lead to jumps between different manifolds, resulting 
in hysteresis for the tip, and ultimately dissipation.

\section{Equilibrium and time-dependent GL simulations} 
To asses the validity of the proposed mechanism, we have performed numerical simulations of the tip-surface interaction
with a full GL free-energy $\FE[\psi(\rr)] =\FE_0+\FE_V$.
For simplicity, a two-dimensional GL functional is considered, since this takes into account the relevant elastic
effects while keeping the simulation simple enough: indeed, the experimental substrate NbSe$_2$~\cite{Langer_NatMat14}
has a quasi-two-dimensional structure, so that volume effects are expected to be not crucial.  
Differently from the experimental system~\cite{McMillan_PRB75}, we will model the CDW as being characterized by a 
single wavevector $\qq$, leading to a simpler order parameter and a clearer effect.
To represent the effect of the tip, the shape of a van der Waals potential $C/r^6$ is integrated over a conical 
tip at distance $d$ from the surface. 
We have found that the result of such a calculation can be reasonably approximated in the main area under the tip 
by a Lorentzian curve:
\begin{equation} \label{eq:ExtPot}
V(\rr;d)=\frac{V_0(d)}{\rr^2+\sigma^2(d)}  \;,
\end{equation}
where $\rr$ is the distance in the plane from the point right below the tip and the parameters are found to scale like 
$V_0(d)=\overline{V}/d$ and $\sigma(d)=\overline{\sigma}d^2$.
Knowing the shape of the perturbation, the total free energy $\FE=\FE_0+\FE_V$ is minimized numerically
on a square grid of points with spacing much smaller than the characteristic wavelength of the CDW, 
imposing a constant boundary condition $\psi_0$ on the sides perpendicular to $\qq$, while setting 
periodic boundary conditions in the other direction to allow for possible phase jumps. 
The minimization is carried out through a standard conjugated gradients algorithm~\cite{NumRec_07}.
The parameters we have employed are order of magnitude estimates of the real parameters, reproducing the relevant 
experimental effects on NbSe$_2$~\cite{Langer_NatMat14} in a qualitative fashion.

\begin{figure}[!tb]\centering
\includegraphics[width=0.48\textwidth]{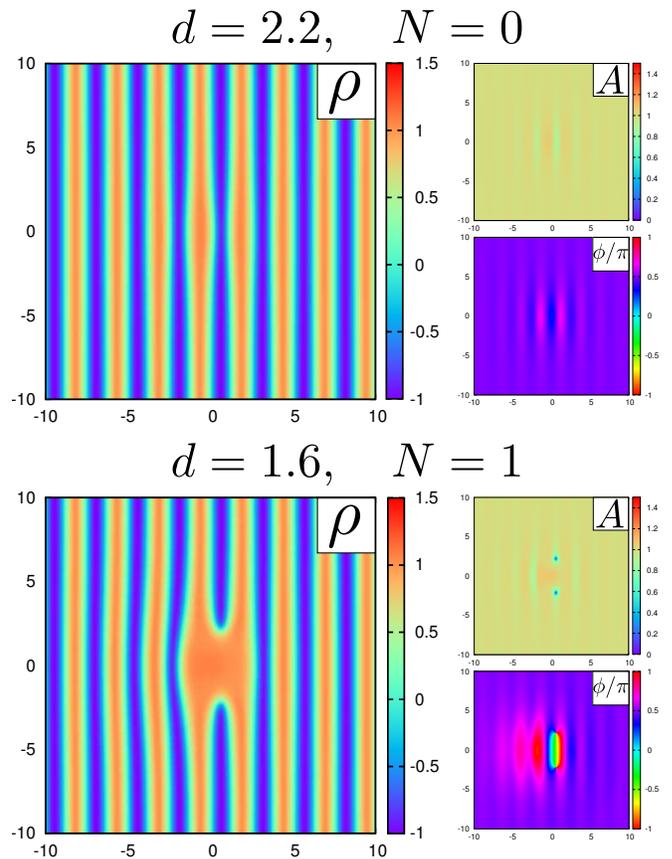}
\caption{\label{fig:minima} Charge density $\Delta \rho$, order parameter amplitude $A$ and potraits of the phase $\phi$  
for minimal free energy solutions with different winding-number $N$ and tip-surface distance $d$ (in nm). Results 
from simulations on a $201\times201$ grid with parameters (see text) $f_0=2$ eV/nm, $\kappa=0.2$ eV, $Q=2.5$ nm, 
$\overline{V}=-9.4$ eV$\cdot$nm, $\overline{\sigma}=1,2$ nm$^{-1}$ and boundary conditions $\psi_0=i$ 
(right and left sides).}
\end{figure}
%
Fig.~\ref{fig:minima} shows the charge density modulation $\Delta\rho$, together with a plot of the  
amplitude $A(\rr)$ and a phase portrait of $\phi(\rr)$, corresponding to GL minima with different winding-number $N$, 
for a non-contact (attractive) tip at different distances $d$. 
The winding-number is calculated along the line passing through the point right below the tip 
(center of the simulation cell) according to Eq.~\eqref{eq:WN}, with $N=0$ being the unperturbed case. 
As predicted, we see upon decreasing $d$ through the first and successive critical distances $d_{01}, d_{12}$, etc.  
the appearance of vortex-antivortex pairs for every unit increase of the winding number. 
These vortices are characterized by a zero of the amplitude $A(\rr)$ and a total change of the phase by $2\pi$ 
on a path around them, as they separate the phase-slippage center from the unaffected area far 
from the tip.

\begin{figure}[!tb]\centering
\includegraphics[width=0.48\textwidth]{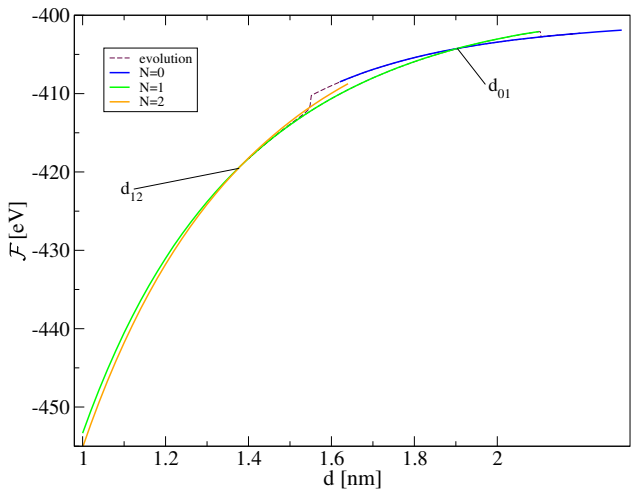}
\caption{\label{fig:encurves} 
Minimal free energy $\FE$ as a function of tip distance $d$ for subspaces with different winding number $N$ 
(full lines) and evolution of $\FE$ during a tip oscillation (dashed line) with $d_0=1.8$ nm, $\bar{d}=0.4$ nm, 
$\omega=6\cdot 10^4$ Hz (other parameters are the same as Fig.~\ref{fig:minima}). 
} 
\end{figure}

Since the solution with a given winding number $N$ represents a local minimum, it is possible to use the minimization 
algorithm, for example  by starting from a reasonable configuration, 
to find solutions in a certain $N$-subspace, even when that is not the global minimum for that given case. 
This allows us to extend the calculation of the local free energy minima in a given $N$ subspace well beyond 
their crossing points, generating a family of free energy curves of definite $N$ as a function of the distance $d$. 
Fig.~\ref{fig:encurves} (full lines) is an example, showing two successive crossing points. We expect each crossing 
to give rise to a first order transition, and thus to a hysteretic peak in the experimental dissipation trace. 
Of course, a more complex CDW configuration or different parameters could give rise to more and different peaks.

\begin{figure}[!b]\centering
\includegraphics[width=0.48\textwidth]{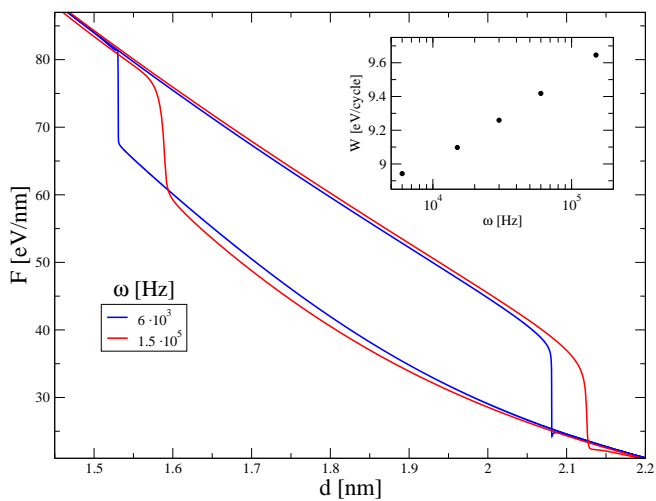}
\caption{\label{fig:fcurves} Force as a function of distance for evolutions with $d_0=1$~nm, $\bar{d}=0.4$~nm and different 
values of $\omega$ with $\Gamma=10^{-7}$~eV$\cdot$s. Inset: total work $W$ as a function of oscillation frequency $\omega$.}
\end{figure}
%
To justify more firmly the validity of the proposed dissipation mechanism, we need to look into the dynamics of the CDW, 
upon varying the tip-surface distance according to the law $d(t)=d_0+\bar{d}\cos(\omega t)$,
to guarantee that the evolution through a crossing point does not lead to immediate relaxation between different $N$-manifolds. 
To do this, the time evolution of the system was simulated, following the time-dependent Ginzburg-Landau equation~\cite{Gorkov_ZETF84}
\begin{equation} \label{eq:TDGL}
-\Gamma\frac{\partial\psi}{\partial t} = \frac{\delta\FE}{\delta\psi^*} \;.
\end{equation}
This equation can be interpreted as an overdamped relaxation of the order parameter towards 
the equilibrium position, with a relaxation rate $\Gamma^{-1}$. 
Integrating this equation (through a standard Runge-Kutta algorithm~\cite{NumRec_07}), the instantaneous force 
$F=-\nabla_d \FE$ as a function of the distance can be computed for a tip performing a 
full oscillation perpendicular to the surface according to the law $d(t)=d_0+\bar{d}\cos(\omega t)$.
Fig.~\ref{fig:fcurves} shows the force evolution during such oscillations at different frequencies. As we can see, the tip
suffers a hysteresis even at low frequencies, since the decay from one manifold to the other happens far from the crossing
point. The area of the loops represents directly the dissipated energy per cycle $W$, as reported in the inset.

\section{Discussion and conclusions} 
We have shown that local surface CDW phase slips and vortex pairs can be introduced by the external
potential of an approaching tip.  In the context of macroscopic CDW conduction noise~\cite{Ong_PRL84,Gruner_PRL81,Gruner_RMP88}, 
the creation and movement of vortices has been invoked earlier in connection with phase slips near the CDW boundaries. 
In a broader context, our system can be placed in between these macroscopic situations and the simple  
models of defect pinning and phase-slip~\cite{Tucker_PRB89} by a localized perturbation.

Experimentally, Langer et al.~\cite{Langer_NatMat14} recently reported AFM dissipation peaks appearing at discrete
tip-surface distances above the CDW material 2H-NbSe$_2$, qualitatively suggesting, in a 1D toy model, 
the injection of $2\pi$ phase slips. The present results describe at the minimal level a theory 
that can explain this type of phenomenon, connecting the phase slip to a vortex-pair formation, and
providing the time dependent portrait of the injection process.

It would be of considerable future interest to explore further this effect in other systems with different characteristics.
In insulating, quasi-one dimensional CDW systems the injected phase slip should also amount to
the injection of a quantized, possibly fractional pairs of opposite charges~\cite{CDW_note2}. 
In a spin density wave system, such as the chromium surface, a nonmagnetic tip would still couple 
to the accompanying CDW~\cite{Kim_PRB13} where surface phase slips could be injected.
In superconductors, the induction of single vortices over Pb thin film islands has been experimentally
verified \cite{Hasegawa_Nanotec10} and the feasibility of controlling single vortices through magnetic force
microscopy (MFM) tips demonstrated \cite{Auslander_NatPhy09}: it would be interesting to probe for
dissipation peaks, as we have addressed above, induced by the MFM tip creation of vortex pairs in thin superconducting films. 

\acknowledgments We acknowledge research support by SNSF, 
through SINERGIA Project CRSII2 136287/1, by ERC Advanced Research Grant N. 320796 MODPHYSFRICT, 
by EU-Japan Project 283214 LEMSUPER, and by MIUR, through PRIN-2010LLKJBX\_001.



\begin{thebibliography}{29}
\expandafter\ifx\csname natexlab\endcsname\relax\def\natexlab#1{#1}\fi
\expandafter\ifx\csname bibnamefont\endcsname\relax
  \def\bibnamefont#1{#1}\fi
\expandafter\ifx\csname bibfnamefont\endcsname\relax
  \def\bibfnamefont#1{#1}\fi
\expandafter\ifx\csname citenamefont\endcsname\relax
  \def\citenamefont#1{#1}\fi
\expandafter\ifx\csname url\endcsname\relax
  \def\url#1{\texttt{#1}}\fi
\expandafter\ifx\csname urlprefix\endcsname\relax\def\urlprefix{URL }\fi
\providecommand{\bibinfo}[2]{#2}
\providecommand{\eprint}[2][]{\url{#2}}

\bibitem[{\citenamefont{Giuliani et~al.}(1976)\citenamefont{Giuliani, Tosatti,
  and Tosi}}]{Giuliani_Tosatti_Tosi_LNC76}
\bibinfo{author}{\bibfnamefont{G.}~\bibnamefont{Giuliani}},
  \bibinfo{author}{\bibfnamefont{E.}~\bibnamefont{Tosatti}}, \bibnamefont{and}
  \bibinfo{author}{\bibfnamefont{M.~P.} \bibnamefont{Tosi}},
  \bibinfo{journal}{Lettere al Nuovo Cimento} \textbf{\bibinfo{volume}{16}},
  \bibinfo{pages}{385} (\bibinfo{year}{1976}).

\bibitem[{\citenamefont{Giuliani and Tosatti}(1978)}]{Giuliani_Tosatti_NC78}
\bibinfo{author}{\bibfnamefont{G.}~\bibnamefont{Giuliani}} \bibnamefont{and}
  \bibinfo{author}{\bibfnamefont{E.}~\bibnamefont{Tosatti}},
  \bibinfo{journal}{Il Nuovo Cimento} \textbf{\bibinfo{volume}{47B}},
  \bibinfo{pages}{135} (\bibinfo{year}{1978}).

\bibitem[{\citenamefont{Giuliani et~al.}(1979)\citenamefont{Giuliani, Tosatti,
  and Tosi}}]{Giuliani_Tosatti_Tosi_JPC79}
\bibinfo{author}{\bibfnamefont{G.}~\bibnamefont{Giuliani}},
  \bibinfo{author}{\bibfnamefont{E.}~\bibnamefont{Tosatti}}, \bibnamefont{and}
  \bibinfo{author}{\bibfnamefont{M.~P.} \bibnamefont{Tosi}},
  \bibinfo{journal}{J. Phys. C - Solid State Phys.}
  \textbf{\bibinfo{volume}{12}}, \bibinfo{pages}{2769} (\bibinfo{year}{1979}).

\bibitem[{\citenamefont{Gr{\"u}ner}(1988)}]{Gruner_RMP88}
\bibinfo{author}{\bibfnamefont{G.}~\bibnamefont{Gr{\"u}ner}},
  \bibinfo{journal}{Rev. Mod. Phys.} \textbf{\bibinfo{volume}{60}},
  \bibinfo{pages}{1129} (\bibinfo{year}{1988}).

\bibitem[{\citenamefont{Overhauser}(1968)}]{Overhauser_PR68}
\bibinfo{author}{\bibfnamefont{A.~W.} \bibnamefont{Overhauser}},
  \bibinfo{journal}{Phys. Rev.} \textbf{\bibinfo{volume}{167}},
  \bibinfo{pages}{691} (\bibinfo{year}{1968}).

\bibitem[{\citenamefont{Peierls}(1955)}]{Peierls_55}
\bibinfo{author}{\bibfnamefont{R.~E.} \bibnamefont{Peierls}},
  \emph{\bibinfo{title}{Quantum Theory of Solids}} (\bibinfo{publisher}{Oxford
  University Press}, \bibinfo{year}{1955}).

\bibitem[{\citenamefont{Woll and Kohn}(1962)}]{Woll_PR62}
\bibinfo{author}{\bibfnamefont{E.~J.} \bibnamefont{Woll}} \bibnamefont{and}
  \bibinfo{author}{\bibfnamefont{W.}~\bibnamefont{Kohn}},
  \bibinfo{journal}{Phys. Rev.} \textbf{\bibinfo{volume}{126}},
  \bibinfo{pages}{1693} (\bibinfo{year}{1962}).

\bibitem[{\citenamefont{Weber et~al.}(2011)\citenamefont{Weber, Rosenkranz,
  Castellan, Osborn, Hott, Heid, Bohnen, Egami, Said, and
  Reznik}}]{Weber_PRL11}
\bibinfo{author}{\bibfnamefont{F.}~\bibnamefont{Weber}},
  \bibinfo{author}{\bibfnamefont{S.}~\bibnamefont{Rosenkranz}},
  \bibinfo{author}{\bibfnamefont{J.-P.} \bibnamefont{Castellan}},
  \bibinfo{author}{\bibfnamefont{R.}~\bibnamefont{Osborn}},
  \bibinfo{author}{\bibfnamefont{R.}~\bibnamefont{Hott}},
  \bibinfo{author}{\bibfnamefont{R.}~\bibnamefont{Heid}},
  \bibinfo{author}{\bibfnamefont{K.-P.} \bibnamefont{Bohnen}},
  \bibinfo{author}{\bibfnamefont{T.}~\bibnamefont{Egami}},
  \bibinfo{author}{\bibfnamefont{A.~H.} \bibnamefont{Said}}, \bibnamefont{and}
  \bibinfo{author}{\bibfnamefont{D.}~\bibnamefont{Reznik}},
  \bibinfo{journal}{Phys. Rev. Lett.} \textbf{\bibinfo{volume}{107}},
  \bibinfo{pages}{107403} (\bibinfo{year}{2011}).

\bibitem[{\citenamefont{Coppersmith}(1990)}]{Coppersmith_PRL90}
\bibinfo{author}{\bibfnamefont{S.~N.} \bibnamefont{Coppersmith}},
  \bibinfo{journal}{Phys. Rev. Lett.} \textbf{\bibinfo{volume}{65}},
  \bibinfo{pages}{1044} (\bibinfo{year}{1990}).

\bibitem[{\citenamefont{Inui et~al.}(1988)\citenamefont{Inui, Hall, Doniach,
  and Zettl}}]{Inui_PRB88}
\bibinfo{author}{\bibfnamefont{M.}~\bibnamefont{Inui}},
  \bibinfo{author}{\bibfnamefont{R.~P.} \bibnamefont{Hall}},
  \bibinfo{author}{\bibfnamefont{S.}~\bibnamefont{Doniach}}, \bibnamefont{and}
  \bibinfo{author}{\bibfnamefont{A.}~\bibnamefont{Zettl}},
  \bibinfo{journal}{Phys. Rev. B} \textbf{\bibinfo{volume}{38}},
  \bibinfo{pages}{13047} (\bibinfo{year}{1988}).

\bibitem[{\citenamefont{Maher et~al.}(1992)\citenamefont{Maher, Adelman,
  Ramakrishna, McCarten, DiCarlo, and Thorne}}]{Maher_PRL92}
\bibinfo{author}{\bibfnamefont{M.~P.} \bibnamefont{Maher}},
  \bibinfo{author}{\bibfnamefont{T.~L.} \bibnamefont{Adelman}},
  \bibinfo{author}{\bibfnamefont{S.}~\bibnamefont{Ramakrishna}},
  \bibinfo{author}{\bibfnamefont{J.~P.} \bibnamefont{McCarten}},
  \bibinfo{author}{\bibfnamefont{D.~A.} \bibnamefont{DiCarlo}},
  \bibnamefont{and} \bibinfo{author}{\bibfnamefont{R.~E.}
  \bibnamefont{Thorne}}, \bibinfo{journal}{Phys. Rev. Lett.}
  \textbf{\bibinfo{volume}{68}}, \bibinfo{pages}{3084} (\bibinfo{year}{1992}).

\bibitem[{\citenamefont{Ong et~al.}(1984)\citenamefont{Ong, Verma, and
  Maki}}]{Ong_PRL84}
\bibinfo{author}{\bibfnamefont{N.~P.} \bibnamefont{Ong}},
  \bibinfo{author}{\bibfnamefont{G.}~\bibnamefont{Verma}}, \bibnamefont{and}
  \bibinfo{author}{\bibfnamefont{K.}~\bibnamefont{Maki}},
  \bibinfo{journal}{Phys. Rev. Lett.} \textbf{\bibinfo{volume}{52}},
  \bibinfo{pages}{663} (\bibinfo{year}{1984}).

\bibitem[{\citenamefont{Gr{\"u}ner et~al.}(1981)\citenamefont{Gr{\"u}ner,
  Zawadowski, and Chaikin}}]{Gruner_PRL81}
\bibinfo{author}{\bibfnamefont{G.}~\bibnamefont{Gr{\"u}ner}},
  \bibinfo{author}{\bibfnamefont{A.}~\bibnamefont{Zawadowski}},
  \bibnamefont{and} \bibinfo{author}{\bibfnamefont{P.~M.}
  \bibnamefont{Chaikin}}, \bibinfo{journal}{Phys. Rev. Lett.}
  \textbf{\bibinfo{volume}{46}}, \bibinfo{pages}{511} (\bibinfo{year}{1981}).

\bibitem[{\citenamefont{Vanossi et~al.}(2013)\citenamefont{Vanossi, Manini,
  Urbakh, Zapperi, and Tosatti}}]{Vanossi_RMP13}
\bibinfo{author}{\bibfnamefont{A.}~\bibnamefont{Vanossi}},
  \bibinfo{author}{\bibfnamefont{N.}~\bibnamefont{Manini}},
  \bibinfo{author}{\bibfnamefont{M.}~\bibnamefont{Urbakh}},
  \bibinfo{author}{\bibfnamefont{S.}~\bibnamefont{Zapperi}}, \bibnamefont{and}
  \bibinfo{author}{\bibfnamefont{E.}~\bibnamefont{Tosatti}},
  \bibinfo{journal}{Rev. Mod. Phys.} \textbf{\bibinfo{volume}{85}},
  \bibinfo{pages}{529} (\bibinfo{year}{2013}).

\bibitem[{\citenamefont{Stipe et~al.}(2001)\citenamefont{Stipe, Mamin, Stowe,
  Kenny, and Rugar}}]{Stipe_PRL01}
\bibinfo{author}{\bibfnamefont{B.~C.} \bibnamefont{Stipe}},
  \bibinfo{author}{\bibfnamefont{H.~J.} \bibnamefont{Mamin}},
  \bibinfo{author}{\bibfnamefont{T.~D.} \bibnamefont{Stowe}},
  \bibinfo{author}{\bibfnamefont{T.~W.} \bibnamefont{Kenny}}, \bibnamefont{and}
  \bibinfo{author}{\bibfnamefont{D.}~\bibnamefont{Rugar}},
  \bibinfo{journal}{Phys. Rev. Lett.} \textbf{\bibinfo{volume}{87}},
  \bibinfo{pages}{096801} (\bibinfo{year}{2001}).

\bibitem[{\citenamefont{Gysin et~al.}(2011)\citenamefont{Gysin, Rast, Kisiel,
  Werle, and Meyer}}]{Gysin_RSI11}
\bibinfo{author}{\bibfnamefont{U.}~\bibnamefont{Gysin}},
  \bibinfo{author}{\bibfnamefont{S.}~\bibnamefont{Rast}},
  \bibinfo{author}{\bibfnamefont{M.}~\bibnamefont{Kisiel}},
  \bibinfo{author}{\bibfnamefont{C.}~\bibnamefont{Werle}}, \bibnamefont{and}
  \bibinfo{author}{\bibfnamefont{E.}~\bibnamefont{Meyer}},
  \bibinfo{journal}{Rev. Sci. Instrum.} \textbf{\bibinfo{volume}{82}},
  \bibinfo{pages}{023705} (\bibinfo{year}{2011}).

\bibitem[{\citenamefont{Langer et~al.}(2014)\citenamefont{Langer, Kisiel,
  Pawlak, Pellegrini, Santoro, Buzio, Gerbi, Balakrishnan, Baratoff, Tosatti
  et~al.}}]{Langer_NatMat14}
\bibinfo{author}{\bibfnamefont{M.}~\bibnamefont{Langer}},
  \bibinfo{author}{\bibfnamefont{M.}~\bibnamefont{Kisiel}},
  \bibinfo{author}{\bibfnamefont{R.}~\bibnamefont{Pawlak}},
  \bibinfo{author}{\bibfnamefont{F.}~\bibnamefont{Pellegrini}},
  \bibinfo{author}{\bibfnamefont{G.~E.} \bibnamefont{Santoro}},
  \bibinfo{author}{\bibfnamefont{R.}~\bibnamefont{Buzio}},
  \bibinfo{author}{\bibfnamefont{A.}~\bibnamefont{Gerbi}},
  \bibinfo{author}{\bibfnamefont{G.}~\bibnamefont{Balakrishnan}},
  \bibinfo{author}{\bibfnamefont{A.}~\bibnamefont{Baratoff}},
  \bibinfo{author}{\bibfnamefont{E.}~\bibnamefont{Tosatti}},
  \bibnamefont{et~al.}, \bibinfo{journal}{Nature Mat.}
  \textbf{\bibinfo{volume}{13}}, \bibinfo{pages}{173} (\bibinfo{year}{2014}).

\bibitem[{\citenamefont{Fukuyama and Lee}(1978)}]{Fukuyama_PRB78}
\bibinfo{author}{\bibfnamefont{H.}~\bibnamefont{Fukuyama}} \bibnamefont{and}
  \bibinfo{author}{\bibfnamefont{P.~A.} \bibnamefont{Lee}},
  \bibinfo{journal}{Phys. Rev. B} \textbf{\bibinfo{volume}{17}},
  \bibinfo{pages}{535} (\bibinfo{year}{1978}).

\bibitem[{\citenamefont{Lee and Rice}(1979)}]{Lee_PRB79}
\bibinfo{author}{\bibfnamefont{P.~A.} \bibnamefont{Lee}} \bibnamefont{and}
  \bibinfo{author}{\bibfnamefont{T.~M.} \bibnamefont{Rice}},
  \bibinfo{journal}{Phys. Rev. B} \textbf{\bibinfo{volume}{19}},
  \bibinfo{pages}{3970} (\bibinfo{year}{1979}).

\bibitem[{\citenamefont{Tucker}(1989)}]{Tucker_PRB89}
\bibinfo{author}{\bibfnamefont{J.~R.} \bibnamefont{Tucker}},
  \bibinfo{journal}{Phys. Rev. B} \textbf{\bibinfo{volume}{40}},
  \bibinfo{pages}{5447} (\bibinfo{year}{1989}).

\bibitem[{\citenamefont{T{\"u}tt{\H{o}} and Zawadowski}(1985)}]{Tutto_PRB85}
\bibinfo{author}{\bibfnamefont{I.}~\bibnamefont{T{\"u}tt{\H{o}}}}
  \bibnamefont{and}
  \bibinfo{author}{\bibfnamefont{A.}~\bibnamefont{Zawadowski}},
  \bibinfo{journal}{Phys. Rev. B} \textbf{\bibinfo{volume}{32}},
  \bibinfo{pages}{2449} (\bibinfo{year}{1985}).

\bibitem[{\citenamefont{Maki}(1995)}]{Maki_PLA95}
\bibinfo{author}{\bibfnamefont{K.}~\bibnamefont{Maki}}, \bibinfo{journal}{Phys.
  Lett. A} \textbf{\bibinfo{volume}{202}}, \bibinfo{pages}{313}
  (\bibinfo{year}{1995}).

\bibitem[{\citenamefont{{Gor'kov}}(1984)}]{Gorkov_ZETF84}
\bibinfo{author}{\bibfnamefont{L.~P.} \bibnamefont{{Gor'kov}}},
  \bibinfo{journal}{Zh. Eksp. Teor. Fiz.} \textbf{\bibinfo{volume}{86}},
  \bibinfo{pages}{1818} (\bibinfo{year}{1984}).

\bibitem[{\citenamefont{McMillan}(1975)}]{McMillan_PRB75}
\bibinfo{author}{\bibfnamefont{W.~L.} \bibnamefont{McMillan}},
  \bibinfo{journal}{Phys. Rev. B} \textbf{\bibinfo{volume}{12}},
  \bibinfo{pages}{1187} (\bibinfo{year}{1975}).

\bibitem[{\citenamefont{Press et~al.}(2007)\citenamefont{Press, Teukolsky,
  Vetterling, and Flannery}}]{NumRec_07}
\bibinfo{author}{\bibfnamefont{W.~H.} \bibnamefont{Press}},
  \bibinfo{author}{\bibfnamefont{S.~A.} \bibnamefont{Teukolsky}},
  \bibinfo{author}{\bibfnamefont{W.~T.} \bibnamefont{Vetterling}},
  \bibnamefont{and} \bibinfo{author}{\bibfnamefont{B.~P.}
  \bibnamefont{Flannery}}, \emph{\bibinfo{title}{Numerical Recipes: The Art of
  Scientific Computing (3rd ed.)}} (\bibinfo{publisher}{Cambridge University
  Press}, \bibinfo{year}{2007}).

\bibitem[{CDW()}]{CDW_note2}
\bibinfo{note}{This feature is not expected in NbSe$_2$, which is metallic and
  in reality an anharmonicity-driven PLD~\cite{Weber_PRL11}. Better candidates
  for these effects would be NbSe$_3$ or TaS$_3$}.

\bibitem[{\citenamefont{Kim et~al.}(2013)\citenamefont{Kim, Logan, Shpyrko,
  Littlewood, and Isaacs}}]{Kim_PRB13}
\bibinfo{author}{\bibfnamefont{H.~C.} \bibnamefont{Kim}},
  \bibinfo{author}{\bibfnamefont{J.~M.} \bibnamefont{Logan}},
  \bibinfo{author}{\bibfnamefont{O.~G.} \bibnamefont{Shpyrko}},
  \bibinfo{author}{\bibfnamefont{P.~B.} \bibnamefont{Littlewood}},
  \bibnamefont{and} \bibinfo{author}{\bibfnamefont{E.~D.}
  \bibnamefont{Isaacs}}, \bibinfo{journal}{Phys. Rev. B}
  \textbf{\bibinfo{volume}{88}}, \bibinfo{pages}{140101(R)}
  (\bibinfo{year}{2013}).

\bibitem[{\citenamefont{Nishio et~al.}(2010)\citenamefont{Nishio, Lin, An,
  Eguchi, and Hasegawa}}]{Hasegawa_Nanotec10}
\bibinfo{author}{\bibfnamefont{T.}~\bibnamefont{Nishio}},
  \bibinfo{author}{\bibfnamefont{S.}~\bibnamefont{Lin}},
  \bibinfo{author}{\bibfnamefont{T.}~\bibnamefont{An}},
  \bibinfo{author}{\bibfnamefont{T.}~\bibnamefont{Eguchi}}, \bibnamefont{and}
  \bibinfo{author}{\bibfnamefont{Y.}~\bibnamefont{Hasegawa}},
  \bibinfo{journal}{Nanotechnology} \textbf{\bibinfo{volume}{21}},
  \bibinfo{pages}{465704} (\bibinfo{year}{2010}).

\bibitem[{\citenamefont{Auslaender et~al.}(2009)\citenamefont{Auslaender, Luan,
  Straver, Hoffman, Koshnick, Zeldov, Bonn, Liang, Hardy, and
  Moler}}]{Auslander_NatPhy09}
\bibinfo{author}{\bibfnamefont{O.~M.} \bibnamefont{Auslaender}},
  \bibinfo{author}{\bibfnamefont{L.}~\bibnamefont{Luan}},
  \bibinfo{author}{\bibfnamefont{E.~W.~J.} \bibnamefont{Straver}},
  \bibinfo{author}{\bibfnamefont{J.~E.} \bibnamefont{Hoffman}},
  \bibinfo{author}{\bibfnamefont{N.~C.} \bibnamefont{Koshnick}},
  \bibinfo{author}{\bibfnamefont{E.}~\bibnamefont{Zeldov}},
  \bibinfo{author}{\bibfnamefont{D.~A.} \bibnamefont{Bonn}},
  \bibinfo{author}{\bibfnamefont{R.}~\bibnamefont{Liang}},
  \bibinfo{author}{\bibfnamefont{W.~N.} \bibnamefont{Hardy}}, \bibnamefont{and}
  \bibinfo{author}{\bibfnamefont{K.~A.} \bibnamefont{Moler}},
  \bibinfo{journal}{Nature Phys.} \textbf{\bibinfo{volume}{5}},
  \bibinfo{pages}{35} (\bibinfo{year}{2009}).

\end{thebibliography}

\end{document}